\documentclass[10pt]{article}

\usepackage{amsthm,amssymb,amsmath,stmaryrd}
\usepackage{graphics}
\usepackage{bbm}
\usepackage{tikz}
\usepackage[plainpages=false,pdfpagelabels=false,colorlinks=true,citecolor=blue,hypertexnames=false]{hyperref}
\usepackage{color}

\usepackage{dsfont}
\usepackage{epsfig}
\usepackage{mathrsfs}
\usepackage{amsmath}
\usepackage{cases}

\newtheorem{thm}{Theorem}[section]

\newtheorem{lem}[thm]{Lemma}
\newtheorem{prop}[thm]{Proposition}
\newtheorem{defn}[thm]{Definition}

\newtheorem{rem}[thm]{Remark}
\newtheorem{exam}[thm]{Example}

\newtheorem{problem}[thm]{Problem}

\makeatletter \@addtoreset{equation}{section}

\def\pf{\noindent {\it Proof.\ }}
\def\qed{\hfill \rule{4pt}{7pt}}

\def\si{\sigma}
\def\la{\langle}
\def\ra{\rangle}

\newcommand{\cR} { {\mathcal{R}}}
\newcommand{\bN} { {\mathbb{N}}}

\newcommand{\bQ} { {\mathbb{Q}}}
\newcommand{\bZ} { {\mathbb{Z}}}

\newcommand{\bF} { {\mathbb{F}}}
\newcommand{\bK} { {\mathbb{K}}}
\newcommand{\bE} { {\mathbb{E}}}

\newcommand{\cO}{ {\mathcal O}}
\begin{document}

\title{Existence Problem of Telescopers: Beyond the Bivariate Case
\thanks{S.\ Chen was supported by the NSFC grants 11501552, 11371143 and
by the President Fund of the Academy of
Mathematics and Systems Science, CAS (2014-cjrwlzx-chshsh). This work was
also supported by the Fields Institute's 2015 Thematic Program on Computer
Algebra in Toronto, Canada. Q.\ Hou and R.\ Wang were supported by the 973 Project, the PCSIRT Project of
the Ministry of Education and the National Science Foundation of China.}}

\author{  \bigskip
Shaoshi Chen$^{1,2}$,  Qinghu Hou$^{3}$,  George Labahn$^2$, Ronghua Wang$^{4}$ \\
$^1$KLMM,\, AMSS, \,Chinese Academy of Sciences, \\Beijing, 100190, China\\
$^2$Symbolic Computation Group,  University of Waterloo, \\Ontario, N2L3G1, Canada\\
$^3$Center for Applied Mathematics, Tianjin University,\\ Tianjin, 300072, China\\
$^4$Center for Combinatorics, Nankai University, \\    \bigskip  Tianjin, 300071, China\\
{\sf schen@amss.ac.cn, qh\_hou@tju.edu.cn}\\
{\sf glabahn@uwaterloo.ca}\\ {\sf wangwang@mail.nankai.edu.cn}
}

\maketitle

\begin{abstract}
In this paper, we solve the existence problem of telescopers for
rational functions in three discrete variables. We reduce the problem
to that of deciding the summability of bivariate rational functions, which
has been solved recently. The existence criteria we present
is needed for detecting the termination of Zeilberger's algorithm to
the function classes studied in this paper.
\end{abstract}

\section{Introduction}\label{SECT:intro}
The method of creative telescoping is an algorithmic tool in the
symbolic evaluation of parameterized definite sums and integrals.
In order to evaluate a multiple sum of a given summand~$f(x, y_1, \ldots, y_n)$ with respect to~$y_1, \ldots, y_n$
with~$x$ a discrete parameter, the key step of creative telescoping is to find
a nonzero linear recurrence operator~$L$ in~$x$ such that
\[
L(f) = \Delta_{y_1} (g_1) + \cdots + \Delta_{y_n}(g_n),
\]
where~$\Delta_{y_i}$ denotes the difference operator in~$y_i$ and the~$g_i$'s
belong to the same class of functions as~$f$. The operator $L$ is then called a \emph{telescoper} for~$f$.
In order to be useful in applications one needs to address two problems: (1) determine whether such an operator~$L$
exists for a given~$f$ and, (2) if a telescoper exists then design an algorithm for computing it.
In this paper we focus on the problem of existence of a telescoper for a given $f$.

The existence of telescopers is closely related to the termination of Zeilberger's algorithm for computing telescopers. Since the 1990's, extensive work has been done around the existence problem. A sufficient condition was first given by Zeilberger~\cite{Zeilberger1990} where it was shown that telescopers exist for all holonomic functions. Later Wilf and Zeilberger in~\cite{Wilf1992}, using a linear algebra approach proved that telescopers
always exist for proper hypergeometric terms. However, holonomicity and properness are
only sufficient conditions, that is, there are cases in which the input functions are not holonomic (proper) but telescopers still exist, see~\cite{Chyzak2009}. The first necessary and sufficient conditions
for the existence of telescopers was given by Abramov and Le \cite{AbramovLe2002}
for rational functions in two discrete variables. This was later extended to the
hypergeometric case by Abramov~\cite{Abramov2003} and to the $q$-hypergeometric case
by Chen et al. in~\cite{ChenHouMu2005}. Recently, the remaining six cases of the
existence problem of telescopers for bivariate mixed hypergeometric terms
are solved in~\cite{Chen2015}.
To our knowledge, all of the previous works were only focusing on the problem for
bivariate functions of a special class. Our long-term goal is to determine necessary and sufficient conditions for the existence problem for general multivariate functions. In this paper, we solve the problem for the starting case, that is,
the case of rational functions in three discrete variables.

The previous existence criteria are all based on reduction algorithms which decompose an input function into the sum of
{a summable function and a non-summable one}. The existence is then detected by checking whether
the non-summable part is of a special form (so-called proper terms).
The reduction algorithms can also be used to decide the summabilty
of univariate functions. Recently, the reduction
algorithms for univariate rational functions were extended to the bivariate
case in~\cite{ChenSinger2014, HouWang2015}. The generalized reduction
is also the main ingredient for the existence problem for rational functions of three
variables. However, the existence problem in the trivariate case is considerably more involved. As an example the rational function~$1/(x+y+z^2)$ is not proper
(even after the reduction), but it does have a telescoper (see Example~\ref{EXam:intro}), a phenomenon which does not happen in
the bivariate case.

The remainder of this paper is organized as follows. The basic notation and concepts on telescopers
are given in Section~\ref{SECT:preliminaries}. In Sections~\ref{SECT:sum} and~\ref{SECT:exponent}, we review the previous
work on solving the summability problem for bivariate rational functions and present special properties of linear recurrence operators.  The existence
problem for general rational functions are reduced to one with simpler rational functions
in Section~\ref{SECT:reduce} with the existence criteria for these special rational functions presented
in Section~\ref{SECT:criteria}. The paper ends with a conclusion along with topics for future research.


\section{preliminaries}\label{SECT:preliminaries}

Let~$\bK$ be a field of characteristic zero and let ${\bE = \bK(x, y, z)}$
be the field of rational functions in~$x$, $y$, $z$ over~$\bK$.
For $f \in \bE$ define the shift operators~$\si_x, \si_y, \si_z$ on~$\bE$ by $\si_x(f)=f(x+1, y, z)$,
$\si_y(f)=f(x, y+1, z)$, and $\si_z(f)=f(x, y, z+1)$, respectively.
Let~$\cR := \bE[S_x, S_y, S_z]$ denote the ring of
linear recurrence operators over~$\bE$, in which~$S_x, S_y, S_z$ commute and~$S_v \cdot f = \si_v(f) \cdot
 S_v$ for any~$f\in \bE$ and~$v\in \{x, y, z\}$.
The action of an operator~$P = \sum_{i, j, k} p_{i, j,k}S_x^iS_y^jS_z^k$ in~$\cR$ on a rational function~$f\in \bE$
is then given by
\[P(f) = \sum_{i, j, k} p_{i, j,k}f(x+i, y+j, z+k). \]
The \emph{difference operators} $\Delta_x,\Delta_y$ and $\Delta_z$
with respect to $x,y$ and $z$ are defined by
\[
\Delta_x = S_x-1,\ \Delta_y =S_y-1,\
\hbox{ and } \, \Delta_z =S_z-1.
\]

A rational function~$f\in \bE$ is said to be \emph{$(\si_y, \si_z)$-summable} 
in~$\bE$ if
$f = \Delta_y(g) + \Delta_z(h)$ for some~$g, h \in \bE$. We also just say summable if the meaning is clear. For brevity, we sometimes just write $f \equiv_{y, z} 0$
if~$f$ is $(\si_y, \si_z)$-summable.

\begin{defn}\label{DEF:tele}
A nonzero linear recurrence operator~$L\in \bK(x)[S_x]$ is called a \emph{telescoper}
for a rational function~$f\in \bE$ if $L(f)$ is~$(\si_y, \si_z)$-summable in~$\bE$, that is,
there exists~$g, h \in \bE$ such that
\[
L(f) = \Delta_y(g) + \Delta_z(h).
\]
\end{defn}
{Then the central problem to be solved in this paper is:
\begin{problem}\label{PROB:tele}
Given~$f\in \bE$, decide whether $f$ has a telescoper in $\bK(x)[S_x]$.
\end{problem}
}

An operator $L\in \bK(x)[S_x]$ is called a \emph{common left multiple} of~$L_1, \ldots, L_m\in \bK(x)[S_x]$
if there exist operators $L_1', \ldots, L_m' \in \bK(x)[S_x]$ such that
\[L = L_1'L_1 = \cdots = L_m' L_m.\]
Since~$\bK(x)[S_x]$ is a left Euclidean domain, such an $L$ always exists.
Amongst all of them, the one of smallest degree in~$S_x$ is called the least common left multiple~(LCLM).
When the field~$\bK$ is computable, e.g., $\bK = \bQ$, then many efficient algorithms for computing LCLM
have been developed~\cite{BronsteinPetkovsek1996, AbramovLeLi2005}.
\begin{rem}\label{REM:sum}
Let~$f = f_1 + \cdots + f_m$ with all~$f_i\in \bE$. If each~$f_i$ has a telescoper~$L_i$
for~$i=1, \ldots, m$, then the LCLM of the $L_i$ is a telescoper for~$f$. This fact follows from the definition of LCLM
along with the commutativity between operators in~$\bK(x)[S_x]$ and the difference operators~$\Delta_y, \Delta_z$.
\end{rem}

Let~$G = \langle \si_x, \si_y, \si_z\rangle$ be the free Abelian group generated by~$\si_x, \si_y, \si_z$.
Let~$f\in \bE$ and $H$ be a subgroup of~$G$. We call $$[f]_H := \{\si(f)\mid \si\in H\}$$ the $H$-orbit at~$f$.
Two elements~$f, g\in \bE$ are said to be $H$-equivalent if~$[f]_H = [g]_H$, denoted by $f \sim_H g$. The relation
$\sim_H$ is an equivalence relation. Typically, we will take~$H = G$ or~$H=\la\si_y, \si_z\ra$ in the rest of this paper.

\begin{exam} Let~$f = y^2+x+2z$ and~$g = y^2+x-4y+2z+7$.
Then~$f$ and~$g$ are $G$-equivalent since~$g = \si_x \si_y^{-2}\si_z(f)$. However they are
not $\la \si_y, \si_z \ra$-equivalent. Indeed, if $g = \si_y^n \si_z^k(f)$ for some~$n, k\in \bZ$ then equating the coefficients
leads to the linear system $\{2n=-4, n^2+2k = 7\}$. But this implies that $n=-1$ and~$k = 3/2$, a contradiction.
\end{exam}

\section{Summability} \label{SECT:sum}

The first necessary step for solving the existence problem of telescopers
is 
to decide whether a given multivariate function~$f(x_1, \ldots, x_n)$
in a specific class of functions is equal to
$\Delta_{x_1}(g_1) + \cdots + \Delta_{x_n}(g_n)$
for some $g_1, \ldots, g_n$ in the same class as~$f$.
For univariate rational functions the summability problem
was first solved by Abramov~\cite{Abramov1975, Abramov1995b},
with alternative methods later presented in~\cite{Paule1995b, Pirastu1995a}.
The Gosper algorithm~\cite{Gosper1978} solves the problem for univariate hypergeometric terms.
This was then used by Zeilberger~\cite{Zeilberger1990c} to design a fast algorithm to construct telescopers
for bivariate hypergeometric terms. The Gosper algorithm was extended further to the $D$-finite case by
Abramov and van Hoeij in~\cite{AbramovHoeij1997, AbramovHoeij1999}, and to more general difference-field setting
by Karr~\cite{Karr1981, Karr1985} and Schneider~\cite{Schneider2004}.
A significant step in the path towards the multivariate case was taken
by Chen et al.\ in~\cite{ChenHouMu2006}, which gave some necessary conditions
for the summability of bivariate hypergeometric terms. Chen and Singer in~\cite{ChenSinger2014} then presented the
first necessary and sufficient condition for the summability of bivariate rational functions.
Based on the theoretical criterion in~\cite{ChenSinger2014}, Hou and Wang~\cite{HouWang2015} then gave a practical algorithm
for deciding the summability in the bivariate rational case.

In this section, we will recall the summability criterion for bivariate rational functions from~\cite{HouWang2015}.
Let~$\bF := \bK(x)$ and~$f \in \bF(y, z)$.
The key idea is to decompose~$f$ into the following form
\[
f = \Delta_y(g) + \Delta_z(h) + r,
\]
where~$g, h\in \bF(y, z)$ and~$r$ is of the form
\begin{equation} \label{EQ:r1}
r = \sum_{i=1}^n\sum_{j=1}^{m_i} \frac{a_{i,j}}{d_i^j}
\end{equation}
with~$a_{i,j}\in \bF(y)[z]$, $\deg_z(a_{i, j})< \deg_z(d_i)$, $d_i\in \bF[y, z]$
are irreducible polynomials, and $d_i$, $d_{i'}$ are not $\la \si_y, \si_z\ra$-equivalent
for any~$i\neq i'$. The existence of such decompositions has been shown in~\cite[Lemma 3.1]{HouWang2015}.
Then $f$ is $(\si_y, \si_z)$-summable if and only if~$r$ is $(\si_y, \si_z)$-summable.
Since shift operators preserve the multiplicities of the fractions~$a_{i, j}/d_i^j$, we have
$r$ is $(\si_y, \si_z)$-summable if and only if $\sum_{i=1}^m a_{i, j}/d_i^j$ is $(\si_y, \si_z)$-summable
for each~$j$. Furthermore, Lemma 3.2 in~\cite{HouWang2015} shows that $\sum_{i=1}^n a_{i, j}/d_i^j$
is $(\si_y, \si_z)$-summable if and only if $a_{i, j}/d_i^j$ is $(\si_y, \si_z)$-summable for all~$i$ with~$1\leq i \leq n$.
After this, the summability problem for general rational functions in~$\bF(y, z)$ is reduced to the summable
problem for simple fractions of the special form~$a/d^j$. The following theorem~\cite[Theorem 3.3]{HouWang2015} then gives a criterion for
deciding the summability of such special fractions.
\begin{thm} \label{THM:sum}
Let~$f=a/d^j\in \bF(y, z)$ with~$d\in \bF[y, z]$ being irreducible, $a\in \bF(y)[z]\setminus \{0\}$ and  $\deg_z(a)<\deg_z(d)$.
Then $f$ is $(\sigma_y,\sigma_z)$-summable if and only if
\begin{itemize}
\item[{\rm (1)}]
there exist integers $t,\ell$ with $t \not= 0$ such that
\begin{equation}\label{c1}
  \sigma_y^{t}(d)=\sigma_z^{\ell}(d),
\end{equation}
\item[{\rm (2)}]
for the smallest positive integer $t$ such that \eqref{c1} holds, we have
$
a=\sigma_y^{t}\sigma_z^{-\ell}(p)-p
$
for some $p\in \bF(y)[z]$ with $\deg_z(p)<\deg_z(d)$.
\end{itemize}
\end{thm}

\begin{exam}
Let $f=1/(y^n+z^n)$ for $n\in\mathbb{N}$. When $n=1$, Theorem~\ref{THM:sum} implies that $f$ must be $(\sigma_y,\sigma_z)$-summable. In fact, we have
$$
\frac{1}{y+z}=\Delta_y\left(\frac{y}{y+z}\right)+\Delta_z\left(\frac{-y-1}{y+z}\right).
$$
However, when $n>1$ there exists no $(t,\ell)\in\mathbb{Z}^2$ such that $t\neq 0$ and $ \sigma_y^{t}(y^n+z^n)=\sigma_z^{\ell}(y^n+z^n)$. Thus in this case $f$ is not $(\sigma_y,\sigma_z)$-summable.
\end{exam}

\begin{defn}\label{DEF:add}
For a rational function~$f\in \bF(y, z)$, we call the triple $(g, h, r)\in \bF(y, z)^3$ an \emph{additive decomposition}
of~$f$ with respect to~$y$ and~$z$ if $f = \Delta_y(g) + \Delta_z(h) + r$, where~$r$ is of the form~\eqref{EQ:r1} and
all fractions~$a_{i, j}/d_i^j$ are not $(\si_y, \si_z)$-summable.
\end{defn}
\begin{rem}
From the decision procedure for summability given above, additive decompositions always exist
for rational functions in~$\bF(y, z)$. However, we remark that such decompositions may not be unique.
\end{rem}

\section{Exponent Separation}\label{SECT:exponent}
In this section, we will present some special properties of linear recurrence operators having to do with separating exponents. This separation of exponents of an operator will be used in next section for separating orbits of shift operators and will help in simplifying the existence problem.

Let~$m\in \bN$ and~$L$ be a nonzero operator in~$\bK(x)[S_x]$. We can always decompose $L$ into the form
\begin{equation}\label{decompose}
 L = L_0 + L_1 + \cdots +L_{m-1},
 \end{equation}
where~$L_i = \sum_{j=0}^{r_i} \ell_{i,j}S_x^{jm+i}$ for $i=0, 1, \ldots, m-1$.
We call such a decomposition an \emph{$m$-exponent separation} of~$L$.  It is clear that~$L=0$
if and only if~$L_i=0$ for all $i$. Denote
\begin{equation}\label{matrixL}
\mathcal{L}_m =
\begin{bmatrix}
L_0 & L_{m-1} & L_{m-2} & \ldots & L_1\\
L_1 & L_0 & L_{m-1} & \ldots & L_2\\
L_2 & L_1 & L_0 & \ldots & L_3\\
\vdots & \vdots & \vdots & & \vdots\\
L_{m-1} & L_{m-2} & L_{m-3}& \ldots  & L_0
\end{bmatrix}.\end{equation}
The next lemma and proposition will show that the $m$ rows of~$\mathcal{L}$ are
linearly independent over the ring~$\bK(x)[S_x]$.
\begin{lem}\label{LM:L}
Suppose
\begin{equation}\label{identity}
[T_0, \ldots , T_{m-1}] \cdot \mathcal{L}_m = 0
\end{equation}
with each $T_k \in \bK(x)[S_x]$. Then $T_0 + \cdots + T_{m-1} = 0$.
\end{lem}
\begin{proof}
Note that
$
\mathcal{L}_m \cdot [1, \ldots, 1]^T = [L, \ldots , L]^T$. Hence any solution of (\ref{identity})
implies that
$$
(T_0 + \cdots + T_{m-1} ) \cdot L = 0.
$$
Since $L$ is nonzero and  $\bK(x)[S_x]$ is a left Euclidean domain we have
$T_0 + \cdots + T_{m-1} = 0$.\
\end{proof}

In fact our goal is to show that each component $T_k$ of (\ref{identity}) is zero, that is, the left kernel of $\mathcal{L}_m$ is trivial.  In order to do this  we do an
$m$-exponent separation of each $T_k$ and look at the resulting decomposition. Suppose first
that
$$
[T_0, \ldots , T_{m-1} ] \cdot \mathcal{L}_m = [R_0, \ldots , R_{m-1}]
$$
and that for each $k$
\begin{eqnarray*}
T_k & = &  T_{k, 0} + T_{k, 1} + \cdots + T_{k, m-1}\\
R_k & = & R_{k, 0} + R_{k, 1} + \cdots + R_{k, m-1}
\end{eqnarray*}
are the $m$-exponent separations for $T_k$ and $R_k$, respectively. Let
$\mathcal{T}$ and $\mathcal{R}$ be the $m \times m$ matrices defined as
\begin{equation}
\mathcal{T} = \begin{bmatrix}\label{matrixT}
T_{0, 0} & T_{1, m-1}& T_{2, m-2} & \ldots & T_{m-1, 1}\\
T_{0, 1} & T_{1, 0} & T_{2, m-1} & \ldots & T_{m-1, 2}\\
T_{0, 2} & T_{1, 1} & T_{2, 0} & \ldots & T_{m-1, 3}\\
\vdots & \vdots & \vdots & & \vdots\\
T_{0, m-1} & T_{1, m-2} & T_{2, m-3}& \ldots  & T_{m-1, 0}
\end{bmatrix}
\end{equation}
and
\begin{equation*}
\mathcal{R} = \begin{bmatrix}
R_{0, 0} & R_{1, m-1}& R_{2, m-2} & \ldots & R_{m-1, 1}\\
R_{0, 1} & R_{1, 0} & R_{2, m-1} & \ldots & R_{m-1, 2}\\
R_{0, 2} & R_{1, 1} & R_{2, 0} & \ldots & R_{m-1, 3}\\
\vdots & \vdots & \vdots & & \vdots\\
R_{0, m-1} & R_{1, m-2} & R_{2, m-3}& \ldots  & R_{m-1, 0}
\end{bmatrix}.
\end{equation*}
Then it is straightforward to show that
\begin{equation}\label{matrix-reduction}
\mathcal{T} \cdot \mathcal{L}_m = \mathcal{R}.
\end{equation}
\begin{prop}\label{Prop:L}
Suppose
\begin{equation}\label{identity2}
[T_0, \ldots , T_{m-1}] \cdot \mathcal{L}_m = 0
\end{equation}
with each $T_k \in \bK(x)[S_x]$. Then $T_k  = 0$ for each $k$.
\end{prop}
\begin{proof}
From (\ref{matrix-reduction}) and (\ref{identity2}) we have that each $R_k = 0$ and hence also that each $R_{k,j}=0$. Thus $\mathcal{T} \cdot \mathcal{L}_m = 0$ and so for each $j = 1, 2, \ldots , m$ we have
$$
[T_{0,j-1}, \ldots ,  T_{j-1,0},  T_{j,m-1}, \ldots , T_{m-1,j}] \cdot \mathcal{L}_m = 0.
$$
From Lemma \ref{LM:L} we get for each $j$
$$
T_{0,j} + T_{1,j-1} + \cdots + T_{m-1,j-m+1} = 0.
$$
This implies $T_k = 0$ for all $k$.
\end{proof}

We will also later need to use the following:
\begin{prop}\label{Prop:L2}
There is a matrix $\mathcal{M} \in \bK(x)[S_x]^{m \times m}$ such that
\begin{equation}\label{diagonal}
\mathcal{M} \cdot \mathcal{L}_m = \mbox{diagonal}(T_0, T_1, \cdots , T_{m-1})
\end{equation}
with nonzero $T_i \in \bK(x)[S_x]$.
\end{prop}
\begin{proof}
From the definition of~LCLM,we know for any nonzero $A,B\in\mathbb{K}(x)[S_x]$,
there always exist nonzero $A',B'\in\mathbb{K}(x)[S_x]$ such that $A' \cdot A+ B' \cdot B=0$.
Similar to the use of division-free the Gaussian elimination over a Euclidean domain, we can find
$\mathcal{M} \in \bK(x)[S_x]^{m \times m}$ satisfying (\ref{diagonal}) (c.f. \cite{BeckermanChengLabahn}. That each diagonal element is
nonzero follows directly from Proposition \ref{Prop:L} since otherwise there would be a nonzero element of
the right kernel of $\mathcal{L}_m$.
\end{proof}

\section{Reduction to simple fractions}\label{SECT:reduce}

In this section, we will reduce the existence problem of telescopers for
rational functions in~$\bE$ into the same problem but for simpler rational functions.

Let~$f\in \bE$ be nonzero with $f = \Delta_y(g) + \Delta_z(h) + r$ and $(g, h, r)$ be an additive
decomposition of~$f$ with respect to~$y$ and~$z$. Then
$f$ has a telescoper in~$\bK(x)[S_x]$ if and only if~$r$ has a telescoper in~$\bK(x)[S_x]$.
As such, we need only study the existence problem for rational functions of the form~\eqref{EQ:r1}.

For any~$\si\in \la \si_x, \si_y, \si_z \ra$ and~$a, b\in \bE$, we have
\begin{equation}\label{eq:red}
\frac{a}{\si^n(b)} = \si(g) - g + \frac{\si^{-n}(a)}{b},
\end{equation}
where
\[
g = \left\{
  \begin{array}{rl}
    \sum \limits_{i=0}^{n-1} \frac{\si^{i-n}(a)}{\si^i(b)}, & \hbox{if $n\geq 0$;} \\[10pt]
    -\sum \limits_{i=0}^{-n-1} \frac{\si^{i}(a)}{\si^{n+i}(b)}, & \hbox{if $n<0$.}
  \end{array}
\right.
\]
Suppose now that $d_{i'}= \sigma_x^m\sigma_y^n\sigma_z^k d_i$
for some index $i \not= i'$ and $m, n, k\in \bZ$ with~$m\geq 0$.
Applying the formula~\eqref{eq:red} repeatedly yields
\[
\frac{b_{i',j}}{d_{i'}^j}
= \Delta_y(u) + \Delta_z(v) + \frac{\sigma_y^{-n} \sigma_z^{-k} (b_{{i'},j})}{\sigma_x^m d_i^j}
\]
for some~$u, v \in \bE$.  With this reduction, we can always decompose~$r$ of the form~\eqref{EQ:r1}
into the form
\begin{equation}\label{EQ:resform}
r =\sum_{i=1}^I\sum_{j=1}^{J_i}\sum_{\ell=0}^{\ell_{i,j}}
\frac{b_{i,j,\ell}}{\sigma_x^{\ell}d_i^j}
\end{equation}
with $b_{i,j,\ell}\in \bK(x,y)[z], d_i\in \bK[x,y,z]$, $\deg_z(b_{i,j,\ell}){<}\deg_z(d_i)$,
and $d_i$ are irreducible polynomials with~$d_i$ and~$d_{i'}$ being in distinct $\la \si_x, \si_y, \si_z \ra$-orbits
for any $1\leq i\neq i' \leq m$.

Let~$\cO =\{p/q \in \bE \mid \deg_z(p)< \deg_z(q)\}$ and $V_m$ be
the set of all rational functions of the form $\sum_{i=1}^I {a_i}/{b_i^m}$,
where~$a_i, b_i, \in \bK(x, y)[z]$, $\deg_z(a_i) < \deg_z(b_i)$ and~$b_i$'s are distinct
irreducible polynomials in the ring $\bK(x, y)[z]$. By definition, the set $V_m$ forms a subspace of
$\cO$ as vector spaces over~$\bK(x, y)$. By the irreducible partial fraction decomposition,
any $f\in \cO$ can be uniquely decomposed
into $f = f_1 + \cdots + f_n$ with~$f_i \in V_i$ and so~$\cO = \bigoplus_{i=1}^\infty V_i$. The following lemma shows that
the space $V_m$ is invariant under certain linear recurrence operators.

\begin{lem} \label{LM:multi}
Let~$f\in V_m$ and~$P\in \bK(x, y)[S_x, S_y, S_z]$. Then $P(f)\in V_m$.
\end{lem}
\begin{proof}
Let~$f = \sum_{t=1}^n a_t/b_t^m$ and~$P= \sum_{i, j, k} p_{i, j, k} S_x^i S_y^jS_z^k$.
For any $\si=\si_x^i\si_y^j\si_y^k$ with~$i, j, k \in \bZ$, $\si(b)$ is
irreducible and~$\deg_z(\si(a))< \deg_z(\si(b))$. Then all of the simple fractions
$\frac{p_{i, j, k}S_x^iS_y^jS_z^k(a)}{S_x^iS_y^jS_z^k(b)}$ appearing in $P(f)$ are
proper in~$z$ and have irreducible denominators. If some of denominators are the same,
we can simplify them by adding the numerators to get a simple fraction. After this simplification,
we see that $P(f)$ can be written in the same form as $f$, so it is in~$V_m$.
\end{proof}

\begin{lem}\label{LEM:split}
Let $r \in \bE$  be of the form~\eqref{EQ:resform}.
Then $r$ has a telescoper  if and only if the summand
$\sum_{\ell=0}^{\ell_{i,j}}\frac{b_{i,j,\ell}}{\sigma_x^\ell d_i^j}$ has a telescoper for all~$i, j$ with~$1\leq i\leq I$
and $1 \leq j \le J_i$.
\end{lem}
\begin{proof}
From Lemma \ref{LM:multi} we see that any $r$ as in \eqref{EQ:resform} has a telescoper
if and only if $\sum_{i=1}^I\sum_{\ell=0}^{\ell_{i,j}}
\frac{b_{i,j,\ell}}{\sigma_x^{\ell}d_i^j}$ has a telescoper for all different multiplicities~$j$.
Also, from Lemma 3.2 in~\cite{HouWang2015} we have that $\sum_{i=1}^I\sum_{\ell=0}^{\ell_{i,j}}
\frac{b_{i,j,\ell}}{\sigma_x^{\ell}d_i^j}$ has a telescoper if and only if
$\sum_{\ell=0}^{\ell_{i,j}}\frac{b_{i,j,l}}{\sigma_x^\ell d_i^j}$ has a telescoper for all $i$ with~$1\leq i \leq I$.
\end{proof}

At this stage we have reduced the existence of telescopers problem for general rational functions to those having the simple
form
$r = \sum_{\ell=0}^{\ell_{i,j}}\frac{b_{i,j,\ell}}{\sigma_x^{\ell}d_i^j}$.
If~$\sigma_x^{\ell'}d_i=\sigma_x^{\ell}\sigma_y^n\sigma_z^kd_i$ for some $\ell\neq\ell'$ and $n,k\in\bZ$, then applying the
formula~\eqref{eq:red}, we get
\[
\frac{b_{i,j,\ell'}}{\sigma_x^{\ell'}d_i^j}
=\frac{b_{i,j,\ell'}}{\sigma_x^{\ell}\sigma_y^n\sigma_z^kd_i^j}
=\Delta_y(u_{i, j})+\Delta_z(v_{i, j})+\frac{\sigma_y^{-n}\sigma_z^{-k}b_{i,j,\ell'}}{\sigma_x^{\ell}d_i^j}
\]
for some~$u_{i, j}, v_{i, j}\in \bK(x, y, z)$.
Repeating the above transformation gives a decomposition
\[
r=\Delta_y(u)+\Delta_z(v)+\sum_{i=0}^{I'} \frac{b'_i}{\sigma_x^id^j},
\]
where~$u, v\in \bK(x, y, z)$ and $\sigma_x^i(d)$ and $\sigma_x^{i'}(d)$ are not
$\la\sigma_y,\sigma_z\ra$-equivalent for $0\leq i\neq i'\leq I'$.

The following lemma  reduces the existence problem for rational functions 
into one whose denominators have distinct orbits.

\begin{lem}\label{LM:reduce}
Let $$r =\sum_{i=0}^I\frac{b_{i}}{\sigma_{x}^i d^j}
\mbox{ with } ~b_{i}\in \bK(x,y)[z], ~~d\in\bK[x,y,z].$$ Suppose $b_i, d$ are irreducible polynomials,
$\deg_z(b_{i})<\deg_z(d)$ with $\sigma_x^id$ and $\sigma_x^{i'}d$ in
distinct $\la \sigma_y,\sigma_z\ra$-orbits, for~$0\leq i\neq i'\leq I$.
Then $r$ has a telescoper if and only if each simple fraction~$\frac{b_{i}}{\sigma_{x}^i d^j}$ has a telescoper
for~$0\leq i\leq I$.
\end{lem}
\pf Sufficiency follows from Remark~\ref{REM:sum}.
For the other direction assume that $L=\sum_{i=0}^\rho \ell_i S_x^i$ (with $\ell_0 \neq 0$) is a telescoper for $r$.
There are two cases to be considered according to whether there exists
a positive integer $m$ such that $\sigma_x^m d=\sigma_y^n\sigma_z^kd$.

\vspace{0.15in}

{\it Case~$1.$} There is no positive integer $m$ such that
$$\sigma_x^md=\sigma_y^n\sigma_z^kd ~~~ \mbox{  for some }~n,k\in\mathbb{Z}.$$
In this case, $\sigma_x^{i}d$ and $\sigma_x^{i'}d$ are in distinct $\la \sigma_y,\sigma_z\ra$-orbits
for any $i\neq i'$. We claim that $\frac{b_i}{\sigma_x^id^j}$ is $(\sigma_y,\sigma_z)$-summable for~$0\leq i\leq I$. Since
\[
L(r)=\sum_{i=0}^{\rho}\sum_{t=0}^{I}\ell_{i}\sigma_x^i(\frac{b_{t}}{\sigma_x^td^j})
    =\sum_{p=0}^{\rho+I}\sum_{i=0}^p\ell_i\sigma_x^i(\frac{b_{p-i}}{\sigma_x^{p-i}d^j})
\]
is $(\sigma_y,\sigma_z)$-summable, according to Lemma 3.2 in~\cite{HouWang2015}, we get
\begin{equation}\label{EQ:summable}
\sum_{i=0}^p\ell_i\sigma_x^i\left(\frac{b_{p-i}}{\sigma_x^{p-i}d^j}\right)=\Delta_y(u_p)+\Delta_z(v_p)
\end{equation}
for any $0\leq p \leq \rho+I$.

We prove the claim by induction. The result is true for $p=0$ in~\eqref{EQ:summable} since then
$
\frac{b_0}{d^j}=\Delta_y(\frac{u_0}{\ell_0})+\Delta_z(\frac{v_0}{\ell_0}).
$
Suppose we have shown that $\frac{b_i}{\sigma_x^id^j}$ is $(\sigma_y,\sigma_z)$-summable for $i=0,1,\dots,k-1$ with $k\leq I$. Letting $p=k$ in~\eqref{EQ:summable}, we get
$$
\sum_{i=0}^k\ell_i\sigma_x^i\left(\frac{b_{k-i}}{\sigma_x^{k-i}d^j}\right)=\Delta_y(u_k)+\Delta_z(v_k).
$$
As $\frac{b_{k-i}}{\sigma_x^{k-i}d^j}$ is $(\sigma_y,\sigma_z)$-summable for all $1\leq i\leq k$, it is easy to check that
$
\sum_{i=1}^k\ell_i\sigma_x^i(\frac{b_{k-i}}{\sigma_x^{k-i}d^j})
$
is also $(\sigma_y,\sigma_z)$-summable. Thus $\frac{b_k}{\sigma_x^kd^j}$ is $(\sigma_y,\sigma_z)$-summable.

\vspace{0.15in}

{\it Case~$2.$} Suppose $\sigma_x^md=\sigma_y^n\sigma_z^kd$ for $m$ a positive integer  and $n,~k$ some integers.
Let $m_0$ be the smallest such integer and
$
\sigma_x^{m_0}d=\sigma_y^{n_0}\sigma_z^{k_0}d.
$
Since $\sigma_x^{i}d$ and $\sigma_x^{i'}d$ are in distinct $(\sigma_y,\sigma_z)$-orbits, we can assume
$
r=\sum_{i=0}^{m_0-1}\frac{b_i}{\sigma_{x}^{i}d^j}.
$
Suppose the $m_0$-exponent separation of $L$ is
\[
L=L_0+L_1+\cdots+L_{m_0-1}.
\]
According to Lemma~$3.1$ and Lemma~$3.2$ in \cite{HouWang2015}, we have
\begin{numcases}{}
L_0\frac{b_{0}}{d^j}~~~~+L_{m_0-1}\frac{b_{1}}{\sigma_xd^j}+\cdots+
L_1\frac{b_{m_0-1}}{\sigma_x^{m_0-1}d^j} \equiv_{y, z} 0\notag\\
L_1\frac{b_{0}}{d^j}~~~~+~~~~~~ L_0\frac{b_{1}}{\sigma_xd^j}+\cdots+
L_2\frac{b_{m_0-1}}{\sigma_x^{m_0-1}d^j} \equiv_{y, z} 0\notag\\
\hspace{30mm}\cdots\notag\\
L_{m_0-1}\frac{b_{0}}{d^j}+L_{m_0-2}\frac{b_{1}}{\sigma_x d^j}+\cdots+
L_0\frac{b_{m_0-1}}{\sigma_x^{m_0-1}d^j} \equiv_{y, z} 0.\notag
\end{numcases}
If we let
$$\mathcal{V}=\left[~\frac{b_{0}}{d^j},~\frac{b_{1}}{\sigma_xd^j}, ~ \ldots, ~ \frac{b_{m_0-1}}{\sigma_x^{m_0-1}d^j}~\right]$$
then we can write this as
$$
\mathcal{L}_{m_0} \cdot \mathcal{V}^{T} \equiv_{y, z} 0,
$$
with $\mathcal{L}_{m_0}$ from (\ref{matrixL}).
From Proposition \ref{Prop:L2} there exists $T_0, \ldots , T_{m-1}$ and a matrix $\mathcal{M}$ having entries from $\bK(x)[S_x]$ such that
\[
\mathcal{M} \cdot \mathcal{L}_{m_0} = \mbox{ diagonal}(T_0, \ldots , T_{m-1}).
\]
By the commutativity between operators in $\mathbb{K}(x)[\sigma_x]$ and the
difference operators $\Delta_y,\Delta_z$, we know $T_i$ is a
telescoper for $\frac{b_i}{\sigma_x^{i}d^j}$ for $0\leq i \leq m_0-1$.
\qed

\section{Existence criteria}\label{SECT:criteria}

Lemma \ref{LM:reduce} from the previous section implies that the telescoper existence problem for rational functions is reduced
to the case of a rational function of the form
\begin{equation}\label{EQ:f}
f = \frac{b(x, y, z)} {c(x, y)d(x, y, z)^\lambda}
\end{equation}
where~$\lambda\in \bN$, $b, d \in \bK[x, y, z]$ with~$\deg_z(b)<\deg_z(d)$. In this section, we will give
a criterion for deciding the existence of telescopers for rational functions of the above form.
If $b$ and $c$ are not primitive, that is, their contents are not $1$, then
we can write
\[
b=b_0(x)b_1(x,y,z) \quad \hbox{and} \quad c=c_0(x)c_1(x,y),
\]
where $b_1, c_1$ are primitive in~$y, z$.
Similar to the proof of Lemma 7.4 in~\cite{Chen2015},
$\frac{b}{cd^j}$ has a telescoper if and only if~$\frac{b_1}{c_1d^j}$
has a telescoper. As such we can assume in form (\ref{EQ:f}) that $b, c, d$
are all primitive in~$y, z$.


As we did in the proof of Lemma \ref{LM:reduce} we will proceed by case distinction according to whether or not $d$ satisfies the condition that
there exists a positive integer $m$ such that
\begin{equation}{\label{Condition1}}
\sigma_x^md=\sigma_y^n\sigma_z^kd \quad \hbox{ for some }n,k\in\mathbb{Z}.
\end{equation}
We may always assume $m$ is the smallest integer satisfying the above condition.
Let us first consider the case that the condition is not satisfied.
In this case, the existence problem will be reduced to the summability problem.
As the summability problem for bivariate rational functions has been solved
in~\cite{ChenSinger2014, HouWang2015}, the existence problem becomes:

\begin{thm} \label{LM:CR-1}
Let $f= b/(cd^\lambda) \in \bE$ satisfy the same conditions as in~\eqref{EQ:f} but that~$d$ does not satisfy condition~\eqref{Condition1}.
Then $f$ has a telescoper if and only if $f$ is $(\sigma_y,\sigma_z)$-summable.
\end{thm}
\begin{proof} The sufficiency is obvious.
For the necessity,
we assume that $L=\sum_{i=0}^{I} \ell_i S_x^i\in \bK(x)[S_x]$ with $\ell_0, \ell_I\neq0$ is a telescoper for $f$.
Then
\[L(f) = \sum_{i=0}^I \frac{\ell_i\si_x^i(b)}{\si_x^i(c)\si_x^i(d^\lambda)} = \Delta_y(g) + \Delta_z(h)\]
for some~$g, h \in \bE$. Since $\sigma_x^m (d)\neq \sigma_y^n\sigma_z^k(d)$ for any positive integer~$m$ and~$n, k\in \bZ$,  we have
$\sigma_x^i(d)$ and $\sigma_x^{i'}(d)$ are in distinct $(\sigma_y,\sigma_z)$-orbits for any $i\neq i'$.
By Lemma~$3.2$ in \cite{HouWang2015}, the summands $\frac{\ell_i\si_x^i(b)}{\si_x^i(c)\si_x^i(d^\lambda)}$ of $L(f)$ are all
$(\si_y, \si_z)$-summable. In particular, $\ell_0 f$ is $(\si_y, \si_z)$-summable.
As~$\ell_0\in \bK(x)\setminus\{0\}$, $f$ is $(\si_y, \si_z)$-summable.
\end{proof}

The second case where \eqref{Condition1} is satisfied is considerably more involved.
Let~$\overline{\bK}$ be the algebraic closure of~$\bK$.
An irreducible polynomial $q\in \overline{\bK}$ is said to be \emph{integer-linear}
in~$x, y$ and~$z$ over~$\overline{\bK}$ if it is of the form~$\alpha_i x + \beta_jy+\gamma_i z + \delta_i$,
where~$\alpha_i, \beta_i, \gamma_i \in \bZ$ and~$\delta_i\in \overline{\bK}$.
A rational function~$f\in \bE$ is said to be \emph{proper} if it can be written
in the form $f = \frac{p}{\prod_{i=1}^I q_i}$,
where~$p, q_i\in \bK[x, y, z]$ and all $q_i$ are integer-linear
in~$x, y$ and~$z$ over~$\overline{\bK}$. By the fundamental theorem in~\cite[p.\ 590]{Wilf1992},
any proper rational function has a telescoper.

The following lemma describes some necessary conditions for the existence of telescopers.
\begin{lem}\label{LM:necessary}
Let $f=b/(cd^\lambda) \in \bE$ satisfy the same conditions as in~\eqref{EQ:f} and that $d$ satisfies the
condition~\eqref{Condition1}. If one of the following conditions
is also satisfied:
\begin{itemize}
\item[$(i)$] there exist $n_1, n_2, k_1, k_2\in \bZ$ with~$n_1, n_2 >0$ such that~$\si_y^{n_1}(d) = \si_z^{k_1}(d)$
and~$\si_x^{n_2}(c) = \si_y^{k_2}(c)$;
\item[$(ii)$] there exists a positive integer $t$ such that $\si_x^{tm}(c) = \si_y^{tn}(c)$,
\end{itemize}
then $f$ has a telescoper.
\end{lem}
\begin{proof}
Suppose that the polynomials $c$ and $d$ satisfy the conditions~\eqref{Condition1} and~$(i)$.
By Lemma 3 in~\cite{AbramovPetkovsek2002a}, the equalities $\si_x^{n_2}(c) = \si_y^{k_2}(c)$ and
$\si_x^m(d)=\si_y^n\si_z^k(d)$ imply that there exist
$p\in \bK[z]$ and~$q\in \bK[z_1, z_2]$ such that
\[c = p(y+\frac{k_2}{n_2}x)\quad \text{and} \quad d = q(y+\frac{n}{m}x, z+\frac{k}{m}x).\]
Furthermore, the equality $\si_y^{n_1}(d) = \si_z^{k_1}(d)$ implies that
there exists $h \in \bK[z]$ such that
\[d = h (z+\frac{k}{m}x +\frac{k_1}{n_1}(y + \frac{n}{m}x)).\]
Thus both $c$ and $d$ factor into products of integer-linear polynomials
in $x, y$, and~$z$ over $\overline{\bK}$. Therefore $f$ is a proper rational function, and hence
it has a telescoper.

Suppose that $c$ satisfies the condition~$(ii)$. Set
\[L = \sum_{i=0}^{\rho}\ell_{i} S_x^{i t m},\]
where~$\rho\in \bN$ and~$\ell_{i} \in \bK(x)$ are to be determined.
Applying the reduction formula~\eqref{eq:red} yields
\begin{align*}
  L(f) & = \sum_{i=0}^\rho \frac{\ell_i \si_x^{itm}(b)}{\si_x^{itm}(cd^\lambda)} = \sum_{i=0}^\rho \frac{\ell_i \si_x^{itm}(b)}{\si_y^{itn}(c)\si_y^{itn}\si_z^{itk}(d^\lambda)}\\
       & = \Delta_y(u) + \Delta_z(v) + \frac{1}{cd^\lambda} \sum_{i=0}^\rho \ell_i \si_x^{itm}\si_y^{-itn}\si_z^{-itk}(b).
\end{align*}
Note that the degrees of $\si_x^{itm}\si_y^{-itn}\si_z^{-itk}(b)$ in $y$ or $z$ are the same as that of~$b$.
Thus all shifts of~$b$ lie in a finite dimensional linear space over~$\bK(x)$.
If $\rho$ is large enough, then there always exists $\ell_i\in \bK(x)$, not all zero, such that
\[
\sum_{i=0}^{\rho}\ell_i \si_x^{i t m }\sigma_y^{-i t n}
          \sigma_z^{-i t k}(b)=0.
\]
As a result $L = \sum_{i=0}^\rho \ell_i S_x^{itm}$ is a telescoper for $f$.
\end{proof}
\begin{exam} \label{EXam:intro}
Let~$f = 1/d$ with~$d = x+y+z^2$. Since $\si_x(d) = \si_y(d)$ and~$c=1$, $f$ has a telescoper
by Lemma~\ref{LM:necessary}.
\end{exam}

Decompose the rational function $f = \frac{b}{cd^\lambda}$ into the form
\[
f = \frac{1}{d^\lambda} \left( p + \frac{B}{C} + \sum_{i=1}^I \sum_{\ell=1}^{m_i} \frac{b_{i, \ell}}{ c_i^\ell}\right),
\]
where~$p\in \bK(x)[ y, z]$, $B, b_{i, \ell}\in \bK[x, y, z], C, c_i\in \bK[x, y]$ with $\deg_y(B)<\deg_y(C)$,
$\deg_y(b_{i,\ell})<\deg_y(c_i)$,
and all of the irreducible factors of~$C$ satisfy the condition~$(ii)$ as in Lemma~\ref{LM:necessary},
but all $c_i$ do not satisfy this condition. By Lemma~\ref{LM:necessary}, $( p + {B}/{C})/d^\lambda$
has a telescoper and so for the existence problem of telescopers we need only consider
\begin{equation}\label{EQ:r2}
r = \frac{1}{d^\lambda}\sum_{i=1}^I \sum_{\ell=1}^{m_i} \frac{b_{i, \ell}}{ c_i^\ell}.
\end{equation}

From now on, we always assume that~$d$ satisfies the condition~\ref{Condition1}.
As before we consider two distinct cases, in this case according to whether or not $d$ satisfies the condition:
\begin{equation}\label{EQ:d2} \text{
$\si_y^{n_1}(d) = \si_z^{k_1}(d)$ for some~$n_1, k_1\in \bZ$ with~$n_1 >0$.}
\end{equation}


\begin{thm}\label{TM:suff1}
Let~$r\in \bE$ be as in~\eqref{EQ:r2}.
Suppose that $d$ satisfies the condition~\eqref{Condition1}
and there are no integers~$n_1, k_1$ with~$n_1>0$ such that~$\si_y^{n_1}(d) = \si_z^{k_1}(d)$.
Then $r$ has a telescoper if and only if $r=0$.
\end{thm}
\begin{proof}
The sufficiency is clear. For the necessity, we assume
$L=\sum_{i=0}^\rho \ell_iS_x^i\in \bK(x)[S_x]$ with~$\ell_0, \ell_\rho \neq 0$ is a telescoper for~$r$. Let $m$ be
the smallest positive integer such that~$\si_x^m(d) = \si_y^n\si_z^k(d)$ for some~$n, k\in \bZ$.
Then $\si_x^i(d)$ and~$\si_x^j(d)$ are in distinct $\la \si_y, \si_z \ra$-orbits if~$m \nmid (i-j)$.
Let $L = L_0 + \ldots + L_{m-1}$ be the $m$-exponent separation of~$L$.
Since the denominators of~$L_i(r)$ are in distinct $\la \si_y, \si_z \ra$-orbits, Lemma 3.2 in~\cite{HouWang2015}
implies that~$L_i(r)$ is $(\si_y, \si_z)$-summable for all~$i$ with~$0\leq i \leq m-1$.
Then~$L_0 \neq 0$ is a telescoper for~$r$.
Write~$L_0 = \sum_{t=0}^T a_t S_x^{tm}$. Then
\begin{align*}
  L_0(r) & = \sum_{t=0}^T \sum_{i=1}^I \sum_{\ell=1}^{m_i} \frac{a_t \si_x^{tm}(b_{i, \ell})}{ \si_x^{tm}(c_i^\ell) \si_x^{tm}(d^\lambda)} \\
         & = \sum_{t=0}^T \sum_{i=1}^I \sum_{\ell=1}^{m_i} \frac{a_t \si_x^{tm}(b_{i, \ell})}{ \si_x^{tm}(c_i^\ell) \si_y^{tn}\si_z^{tk}(d^\lambda)}\\
         & = \Delta_y(u) + \Delta_z(v) + \frac{h}{d^\lambda}
\end{align*}
where
\[h = \sum_{t=0}^T \sum_{i=1}^I \sum_{\ell=1}^{m_i} \frac{a_t \si_x^{tm}\si_y^{-tn}\si_z^{-tk}(b_{i, \ell})}{ \si_x^{tm}\si_y^{-tn}(c_i^\ell)}.\]
Since $L_0(r)$ is $(\si_y, \si_z)$-summable but $d$ does not satisfy condition~\eqref{EQ:d2},
Theorem~\ref{THM:sum} implies that~$h=0$.
By Lemma~\ref{LM:multi}, for each multiplicity~$\ell$, we have
\[h_\ell = \sum_{t=0}^T \sum_{i=1}^I
\frac{a_t \si_x^{tm}\si_y^{-tn}\si_z^{-tk}(b_{i, \ell})}{ \si_x^{tm}\si_y^{-tn}(c_i^\ell)}=0. \]
We first claim that there exists a polynomial $p \in \Omega :=\{c_i \mid 1\leq i \leq I\}$ such that
$p \neq \si_x^{\nu m}\si_y^{-\nu n}(q)$ for any $q\in \Omega$ and~$\nu\in\mathbb{N}$.
We prove this claim by contradiction. Suppose that for any~$p_1\in \Omega$, there always
exists $p_2 \in \Omega$ such that~$p_1 = \si_x^{\nu_1 m}\si_y^{-\nu_1 n}(p_2)$ for some positive integer~$\nu_1$.
If $p_1 = p_2$, then we get a contraction with the assumption on~$c_i$'s in~\eqref{EQ:r2}.
If~$p_1\neq p_2$, then there exists~$p_3\in \Omega$ such that~$p_2 = \si_x^{\nu_2 m}\si_y^{-\nu_2 n}(p_3)$
for some positive integer~$\nu_2$. Continuing this process, we get a sequence of polynomials $p_1, p_2, \ldots \in \Omega$.
Since~$\Omega$ is a finite set, $p_i = p_j$ for some~$i<j$ in this sequence. Then
$p_i = \si_x^{\nu m}\si_y^{-\nu n}(p_i)$ with~$\nu =\nu_i+\cdots \nu_{j-1}>0$,  a contradiction.
This completes the proof of the claim.

Suppose now that $c_1$ is such an element in $\Omega$ satisfying $c_1 \neq \si_x^{\nu m}\si_y^{-\nu n}(q)$ for any $q\in \Omega$ and~$\nu\in\mathbb{N}$.
Then the fraction $\frac{a_0b_{1, \ell}}{c_1^\ell}$ has a different irreducible denominator
from the other fractions in~$h_\ell$ which implies that $a_0b_{1, \ell} = 0$. Since~$a_0\neq 0$ we have that
$b_{1, \ell} =0$ for all~$\ell$. We can now repeat the argument for the set~$\Omega\setminus\{c_1\}$ to get
$b_{i, \ell}=0$ for all $i=2, \ldots, n$ and all~$\ell$. Thus, $r=0$.
\end{proof}

{
\begin{exam}
Let $$f=\frac{xy+xz+y^2+yz+1}{(x+y)\left((x+y)^2+z^2\right)}.$$ In order to decide whether there exists a telescoper for $f$,  we first rewrite $f$ into $$f=\left(y+z+\frac{1}{x+y}\right)\cdot\frac{1}{(x+y)^2+z^2}.$$ Letting $d=(x+y)^2+z^2$ one has $\sigma_x d=\sigma_y d$ and hence from Remark~\ref{REM:sum} and Lemma~\ref{LM:necessary} we see that $f$ has a telescoper.  In fact, following the proof of Lemma~\ref{LM:necessary}, we can determine that
$$L_1=S_x^2-2S_x+1=(S_x-1)^2 ~~\mbox{ and } ~~ L_2=S_x-1$$
are telescopers for $\frac{y+z}{d}$ and for $\frac{1}{(x+y)d}$, respectively. Thus $L=(S_x-1)^2$ is a telescoper for $f$.
\end{exam}
}

We now study the case when $d$ satisfies the condition~\eqref{EQ:d2}. Assume that $n_1$
is the smallest positive integer such that $\si_y^{n_1}(d) = \si_z^{k_1}(d)$ for some~$k_1\in \bZ$.
By Lemma~\ref{LM:necessary}, all the fractions $\frac{b_{i, \ell}}{c_i^\ell d^\lambda}$ in~\eqref{EQ:r2} with~$c_i$
satisfying the condition: for all~$i$,
\begin{equation}\label{EQ:c}
\text{$\si_x^{n_i}(c_i) =\si_y^{k_i}(c_i)$ for some~$n_i, k_i\in \bZ$ with $n_i>0$}
\end{equation}
have telescopers. It remains to study the existence problem of telescopers for rational functions
of the form
\begin{equation}\label{EQ:r3}
r = \sum_{i=1}^I\sum_{\ell=1}^{m_i} \frac{b_{i,\ell}}{c_i^{\ell}d^\lambda},
\end{equation}
where~$b_{i, \ell}\in \bK[x, y, z], c_i\in \bK[x, y], \deg_y(b_{i, \ell})<\deg_y(c_i)$, where the
$c_i$ are irreducible polynomials such that condition~\eqref{EQ:c} is not satisfied.

\begin{thm}\label{TM:suff2}
Let~$r$ be of the form ~\eqref{EQ:r3} with~$d$ satisfying conditions~\eqref{Condition1} and~\eqref{EQ:d2}
and where~$c_i$'s do not satisfy the condition~\eqref{EQ:c}.
Then $r$ has a telescoper if and only if
$$r_\ell~:=~ \sum_{i=1}^{I}\frac{b_{i,\ell}}{c_i^{\ell}d^\lambda}$$ is $(\sigma_y,\sigma_z)$-summable
for all~$\ell$.
\end{thm}
\begin{proof}
The sufficiency follows from Remark~\ref{REM:sum}. For the necessity, we assume that~$L$ is
a telescoper for~$r$. By the same argument as in the proof of Theorem~\ref{TM:suff1}, we may always
assume that $L = \sum_{t=0}^T a_t S_x^{tm}$ with~$a_0 \neq 0$. The same calculation as in
the proof of Theorem~\ref{TM:suff1} then yields
\[
  L(r) = \Delta_y(u) + \Delta_z(v) + \frac{1}{d^\lambda} h,
\]
where $u, v \in \bK(x, y, z)$ and $h = Q(\sum_{i=1}^I \sum_{\ell=1}^{m_i} b_{i, \ell}/c_i^\ell)$ with
\[Q = \sum_{t=0}^T a_t S_x^{tm}S_y^{-tn}S_z^{-tk} \in \bK(x)[S_x, S_y,S_z].\]
Since $L(r)$ is $(\si_y, \si_z)$-summable but $d$ satisfies the condition~\eqref{EQ:d2},
Theorem~\ref{THM:sum} implies that
$h = \si_y^{n_1}\si_z^{-k_1}(p) -p$,
where~$p\in \bK(x, y)[z]$ with~$\deg_z(p)<\deg_z(d)$.
By Lemma~\ref{LM:multi}, for each multiplicity~$\ell$, we have
\[h_\ell = Q\left(\sum_{i=1}^I \frac{b_{i, \ell}}{c_i^\ell}\right)=\si_y^{n_1}\si_z^{-k_1}(p_\ell) -p_\ell. \]
Let $\triangle :=\{c_i\mid 1\leq i\leq I\}$. As in the argument for the proof of Theorem~\ref{TM:suff1},
we may assume $c_1\in\triangle$ satisfying $c_1\neq \sigma_x^m\sigma_y^n c_i$ for any $c_i\in\triangle$,
when $m, n \in \mathbb{Z}$ with $m>0$. Note that there may exist some $c_i\in\triangle\setminus\{c_1\}$ such that $c_1 = \sigma_y^nc_i$ for some $n\in\bZ$,
and we will let
$$\triangle_1=\{i\mid 1\leq i\leq I, c_i=\sigma_y^nc_1 \hbox{ for some }n\in\mathbb{Z}\}.$$
Continuing now with $\triangle \setminus \triangle_1$,
we will find $c_1, c_2,\ldots,c_M \in\triangle$ and $\triangle_1,\triangle_2,\ldots,\triangle_M$
such that for $1\leq i<i'\leq M$, we have
$
c_i\neq \sigma_x^m\sigma_y^nc_{i'},\hbox{ when }m,n\in\mathbb{Z},\ m>0
$
and
$
\{1,2,\ldots, I\}=\bigcup_{i=1}^{M}\triangle_i.
$
We can therefore rewrite $h_\ell$ as
\begin{equation}\label{EQ:numerator1}
Q\left( \sum_{j=1}^M \sum_{i\in \triangle_j}
\frac{b_{i, \ell}}{c_i^\ell} \right)=\si_y^{n_1}\si_z^{-k_1}(p_\ell) -p_\ell.
\end{equation}
Since $p_{\ell}\in\bQ(x,y)[z]$, we can decompose it into
\[
p_{\ell}=\sum_{j =1}^{M} \sum_{t=\alpha_j}^{\beta_j}
            \frac{u_{j,t}}{\si_y^t(c_j^{\ell})}+q_{\ell},
\]
where $\alpha_i,\beta_i\in\mathbb{Z}$ and $q_{\ell}$ contains no term of the form
$\frac{u_{j,t}}{\si_y^t(c_j^{\ell})}$ in its irreducible partial fraction decomposition
with respect to $y$.
According to Equation~\eqref{EQ:numerator1} and the uniqueness of irreducible partial fraction decomposition along with the fact that $a_0\in \bK(x)\setminus\{0\}$,
we derive that
\[
\sum_{i\in\triangle_1}\frac{b_{i,\ell}}{c_i^{\ell}}
      =\sigma_y^{n_1}\sigma_{z}^{-k_1}(h_{1,\ell})-h_{1,\ell},
\]
where $h_{1,\ell}=\frac{1}{a_0}\sum_{t=\alpha_1}^{\beta_1}
       \frac{u_{1,t}}{\si_y^t(c_1^{\ell})}$.
Collecting all the terms with the denominator $\la\sigma_x,\sigma_y\ra$-equivalent to $c_1$ in Equation~\eqref{EQ:numerator1}, we obtain
\begin{align}\label{EQ:c1}
Q\left(\sum_{i\in \triangle_1}\frac{b_{i, \ell}}{c_i^\ell}\right) &= Q \left(\si_y^{n_{1}}\si_k^{-k_{1}}(h_{1,\ell})-h_{1,\ell}\right)\\
  & = \sigma_y^{n_{1}}\sigma_k^{-k_{1}}\left(p_{1,\ell}\right)-p_{1,\ell} \label{EQ:c1b}
\end{align}
with $p_{1,\ell}=Q(h_{1,\ell})$.
Subtracting Equation~\eqref{EQ:c1b} from Equation~\eqref{EQ:numerator1}, we obtain
\begin{equation}\label{EQ:numerator2}
Q\left( \sum_{j=2}^M \sum_{i\in \triangle_j}
\frac{b_{i, \ell}}{c_i^\ell} \right)=\si_y^{n_1}\si_z^{-k_1}(p_\ell^{\star}) -p_\ell^{\star}
\end{equation}
with $p_{\ell}^{\star}=p_{\ell}-p_{1,\ell}$.
Now we can repeat the arguments for the set~$\Delta\setminus\{\Delta_1\}$ and Equation~\eqref{EQ:numerator2} to get
\[
\sum_{i\in\triangle_j}\frac{b_{i,\ell}}{c_i^{\ell}}=\si_y^{n_{1}}\si_z^{-k_{1}}
            (h_{j,\ell})-h_{j,\ell}
\]
for all $j=1, \ldots, M$ and all~$\ell$.
Then~$\sum_{i\in\triangle_j}\frac{b_{i,\ell}}{c_i^{\ell}d^\lambda}$
is $(\si_y, \si_z)$-summable by Theorem~\ref{THM:sum} and thus $\sum_{i=1}^{I}\frac{b_{i,\ell}}{c_i^{\ell}d^{\lambda}}$ is $(\sigma_y,\sigma_z)$-summable for all~$\ell$.
This completes the proof.
\end{proof}

{
\begin{exam}
Let
$$
f=\frac{x^4+2x^2y^2+y^4+x^3+3yx^2+y^3-xy^2+x^2-xy}
            {(x+y)(x^2+y^2+2y+1)(x^2+y^2)(x+y+z)^2}.
$$
To solve the existence problem of telescopers for $f$, we firstly need to decompose
$$
f=\left(\frac{1}{x+y}+\frac{y+1}{x^2+y^2+2y+1}-\frac{y}{x^2+y^2}\right)
  \cdot\frac{1}{(x+y+z)^2}.
$$
Letting $d=x+y+z$, we have $\sigma_xd=\sigma_yd$ and $\sigma_yd=\sigma_zd$.
As in the proof of Lemma~\ref{LM:necessary}, we get that  $L=S_x-1$ is a telescoper for $\frac{1}{(x+y)(x+y+z)^2}$. Theorem~\ref{THM:sum} then guarantees
$$
\left(\frac{y+1}{x^2+y^2+2y+1}-\frac{y}{x^2+y^2}\right)
  \cdot\frac{1}{(x+y+z)^2}.
$$
is $(\sigma_y,\sigma_z)$-summable, so $L=S_x-1$ is a telescoper for $f$.
\end{exam}
}

\begin{rem}
To test the existence of telescopers for a simple fraction, one first needs to
test the conditions~\eqref{Condition1}, \eqref{EQ:d2} and~\eqref{EQ:c} satisfied by the polynomials $d$ and $c_i$'s. This amounts to solving
the following problem:

\begin{problem}[Integer Shift Equivalence Testing Problem]
Let $\bK$ be any computable field of characteristic zero and
$\si_i$ be the shift operator w.r.t.\ $x_i$ on $\bK[x_1, ..., x_n]$.  Given $p \in \bK[x_1, ..., x_n]$, to decide whether
there exist integers $m_1, ..., m_n$ with
$m_1 >0$ such that  $\si_1^{m_1} \cdots \si_n^{m_n}(p)  = p$.
\end{problem}
This problem is a special case of the problem proposed and solved by
Grigoriev in~\cite{Grigoriev1996,Grigoriev1997} and more recently by Dvir et al. in~\cite{Dvir2014}.
Theorems~\ref{TM:suff1} and~\ref{TM:suff2} reduce the problem to that of testing the summability of bivariate rational functions.
For this, we can apply the algorithm in~\cite{HouWang2015}.
As such the existence problem in this case is solved.
\end{rem}

\section{Conclusion}
In this paper, we solve the existence problem of telescopers for
rational functions in three discrete variables. We give a procedure
which reduces the problem to a special shift equivalence testing problem
and the summability problem of bivariate rational functions.
Those problems have recently been solved.

In terms of future research, the first direction is to solve the existence problem of telescopers for multivariate rational functions
or a more general class of functions, for example, hypergeometric terms. This would include both efficient algorithms and implementations. A crucial first step is solving the
summability problem for these functions. This is also a challenging problem in symbolic summation as noted in~\cite{Andrews1993}.

%
%

\end{document}